%% file: manuscript-R1.tex
\documentclass[journal]{IEEEtran} 
\usepackage[bookmarks=false]{hyperref}
\hypersetup{
    colorlinks,
    citecolor=black,
    filecolor=black,
    linkcolor=black,
    urlcolor=black
}

\ifCLASSINFOpdf
  \usepackage[pdftex]{graphicx}
  \graphicspath{{figures/}{biographies/}}
  \DeclareGraphicsExtensions{.pdf,.jpeg,.png}
\else
  \usepackage[dvips]{graphicx}
  \DeclareGraphicsExtensions{.pdf,.jpeg,.png,.eps}
\fi

\usepackage{amsmath, amsfonts, mathtools, empheq}
\usepackage{amssymb}  
\usepackage[table, dvipsnames]{xcolor}
\usepackage{enumitem}
\usepackage{tabularx}

\usepackage{lipsum,adjustbox}
\usepackage{pbox}
\usepackage{subcaption}

\usepackage{multirow}
\usepackage{longtable}
\usepackage{colortbl}
\usepackage{cite}
\usepackage{url}
\usepackage{hyperref}

\usepackage{balance}
\usepackage{xtab, afterpage}

\usepackage [english]{babel}
\usepackage [autostyle, english = american]{csquotes}
\MakeOuterQuote{"}

\usepackage{bigstrut}
\usepackage{booktabs}
\usepackage{array}
\newcolumntype{L}{>{$}l<{$}}
\newcolumntype{C}{>{$}c<{$}}
\newcolumntype{R}{>{$}r<{$}}

\input{figures/style.tex}

\definecolor{row}{rgb}{0.4, 0.6, 0.8}

\newcommand{\khan}[1]{\textcolor{black}{#1}}

\begin{document}
%
\title{ML-based Handover Prediction and AP Selection in Cognitive Wi-Fi Networks}
%
%
%

\author{
  \IEEEauthorblockN{
  	Muhammad Asif Khan\IEEEauthorrefmark{1},
    Ridha Hamila\IEEEauthorrefmark{2},
    Adel Gastli\IEEEauthorrefmark{2},
    Serkan Kiranyaz\IEEEauthorrefmark{2} and
    Nasser Ahmed Al-Emadi\IEEEauthorrefmark{2}
  }
  
  \IEEEauthorblockA{
  	Department of Electrical Engineering, Qatar University\IEEEauthorrefmark{1}\IEEEauthorrefmark{2}\\
    Email:
        asifk@ieee.org\IEEEauthorrefmark{1} 
        \{hamila, adel.gastli, mkiranyaz, alemadin\}@qu.edu.qa\IEEEauthorrefmark{2} 
  }
  }


\maketitle

\begin{abstract}
Device mobility in dense Wi-Fi networks offers several challenges. Two well-known problems related to device mobility are handover prediction and access point selection. Due to the complex nature of the radio environment, analytical models may not characterize the wireless channel, which makes the solution of these problems very difficult. Recently, cognitive network architectures using sophisticated learning  techniques are increasingly being applied to such problems. In this paper, we propose data-driven machine learning (ML) schemes to efficiently solve these problems in wireless LAN (WLAN) networks. The proposed schemes are evaluated and results are compared with traditional approaches to the aforementioned problems. The results report significant improvement in network performance by applying the proposed schemes. The proposed scheme for handover prediction outperforms traditional methods i.e. \khan{received signal strength} method and traveling distance method by reducing the number of unnecessary handovers by 60\% and 50\% respectively. Similarly, in AP selection, the proposed scheme outperforms the \khan{strongest signal first} and \khan{least loaded first} algorithms by achieving higher throughput gains up to 9.2\% and 8\% respectively.

\end{abstract}


\IEEEpeerreviewmaketitle

\section{Introduction} \label{sec:intro}
Wi-Fi Networks are experiencing two paradigm shifts in terms of size and applications. The size of Wi-Fi networks has increased from small home and office networks to \khan{large-scale} ultra-dense networks, also referred to as \khan{Overlapping Basic Service Set (OBSS)}. At the same time, there has been an increasing number of novel applications and services such as content distribution \cite{khan2019optimal}, Internet of Things (IoT) \cite{Pirayesh2020}, Intelligent Transportation Systems (ITS) \cite{Zeadally2020, omar2016}, Device-to-Device (D2D) based cooperative networking \cite{khan2017wi, cherif2017p2p} and Unmanned Aerial Vehicles (UAVs) \cite{khan2019novel} that are using Wi-Fi as a communication technology. These two paradigm shifts pose several challenges in legacy Wi-Fi networks. Two fundamental challenges in ultra-dense Wi-Fi networks are (i) handover prediction and (ii) access point (AP) selection.

\subsection{Handover Prediction Problem}
\label{subsec:handover_pred}
Handover prediction refers to the problem of anticipating about the connection state of a mobile device associated with an AP. Handover prediction can play a key role in providing seamless connectivity in next generation networks. It brings several potential benefits; \khan{firstly} the accurate prediction of the handover event allows to timely initiate the transfer of connection to a new AP to reduce \khan{the} handover delay. Secondly, it prevents unnecessary handovers (i.e. ping-pongs) to avoid connection disruptions in highly dynamic networks.
\par
\begin{figure}[htbp]
\centering
\includegraphics[width=0.7\columnwidth]{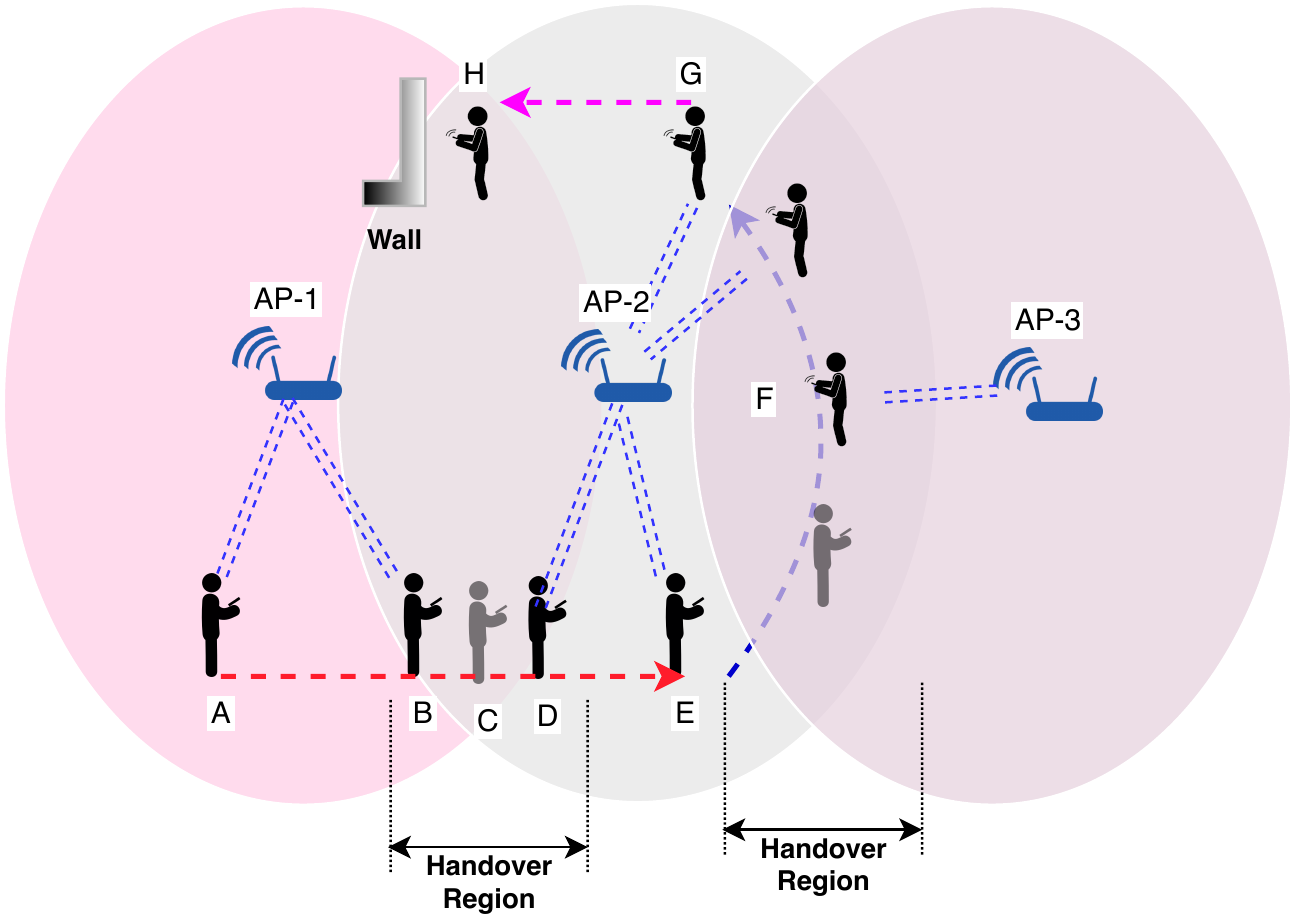}
\caption{Inter-BSS handover scenario.}
\label{fig:handover}
\end{figure}

Handover prediction can be challenging in some cases. Fig. \ref{fig:handover} illustrates different scenarios of inter-BSS \khan{handovers} in Wi-Fi networks. A Wi-Fi user travels from point A to point E (follows the trajectory shown as red, dashed line). Assume that \khan{the user} passes through the region where the radio coverage of AP-1 and AP-2 overlaps, the received signal strength (RSS) drops below the threshold value and it starts scanning for alternate connection. \khan{Meanwhile}, \khan{when the user} moves a bit farther to point D, it discovers AP-2 with a stronger signal. It \khan{disassociates} from AP-1 and associates to AP-2 \khan{($1^{st}$ handover)}. The user continues to move and follows the trajectory from point E to G (dashed blue line) and again passes through an overlapping region of AP-2 and AP-3. At point F, the user changes association to AP-3 \khan{($2^{nd}$ handover)} and back to AP-2 when it moves a little farther \khan{($3^{rd}$ handover)}. The user moves ahead and follows the third trajectory from point G to H, and changes association to AP-1 when it approaches to point H \khan{($4^{th}$ handover)}. At Point H, the user can't move farther towards AP-1 due to hindrance and the signal form AP-2 becomes stronger even with a slight movement in any direction \khan{($5^{th}$ handover)}. 
\par
From the above discussion, it becomes obvious that there are some cases where the handover shall not take place despite the signal strength drops slightly below the threshold level to avoid \khan{the} \textit{ping-pong} effect.

\subsection{Access Point Selection Problem}
\label{subsec:ap_select}
\begin{figure}[htbp]
\centering
i\includegraphics[width=0.7\columnwidth]{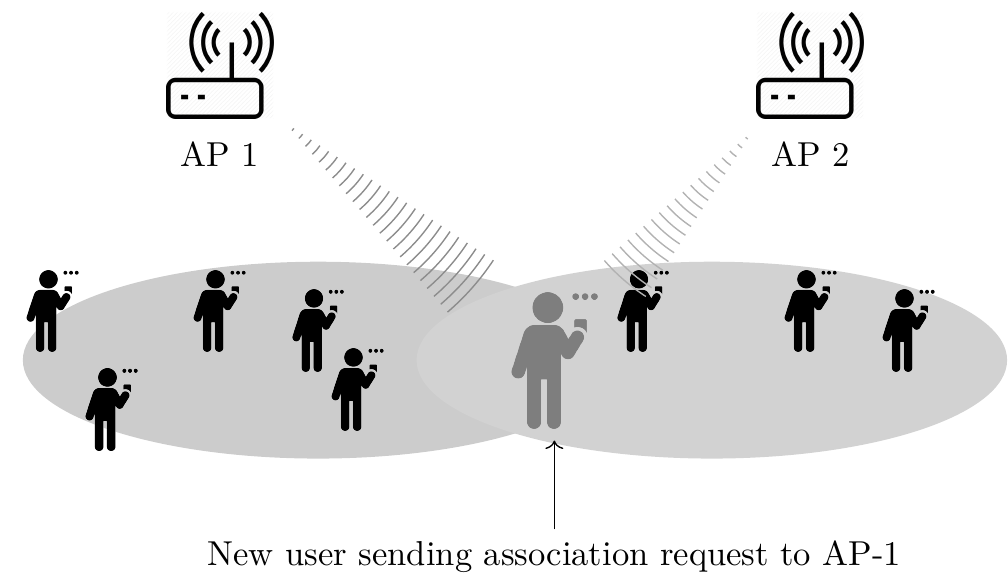}
\caption{User association in overlapping BSS's.}
\label{fig:obss}
\end{figure}

Assuming a Wi-Fi device is located in the transmission range of more than one AP, it can associate with either one as shown in Fig. \ref{fig:obss}. By default, a station associates to the AP from which it first receives a beacon or a probe response frame. However, in practice such kind of automatic association of stations can cause performance degradation \khan{when the connection to the selected AP is weak}. The optimal selection of an access point in dense WLANs is crucial for network performance.
\par
The legacy methods for users association in WLANs are:  (i) Strongest Signal First (SSF) and (ii) Least Loaded First (LLF). Both the SSF and LLF association methods have shortcomings. For instance, in SSF scheme, a station associates to the AP from which it receives a stronger radio signal, however, if the AP is over-utilized, the association of more stations can cause congestion \khan{in the BSS which leads to} increase in the packet loss and the packet end-to-end delay \cite{balachandran_2002, bejerano_2004, mhatre_2006}. On the other hand, in LLF scheme, the selection of the least loaded AP provides load balancing \khan{at multiple APs}, however it may force a station to associate with a distant AP. \khan{Consequently, the} station suffers from poor connection quality. To address these shortcomings of SSF and LLF schemes, the authors in \cite{vasudevan_2005} propose a new metric for AP selection named as \textit{potential bandwidth} which is defined as, "the MAC layer bandwidth that an end-host is likely to receive if it were to affiliate with a given access point". The new metric takes into account the signal strength as well as the AP load and additionally the contention on the wireless medium.  However, the technique in \cite{vasudevan_2005} may not achieve the desired performance if the APs uses different beacon frequencies. It is therefore necessary to devise an AP selection strategy that improves the overall network performance while meeting the demands of \khan{new users}.
\par


Recently, new architectures based on Software Defined Network (SDN) and Cognitive Networking (CN) paradigms are being proposed in the literature \cite{ali_2013, danieletto_2014, slavik_2008}. SDN \cite{kim_2013, afsane_2017, Malbasic2020} refers to the type of networks in which the control and data forwarding functions are separated. In these architectures, the network devices such as switches, routers and access points act as non-intelligent data forwarding devices while the intelligent functions such as data routing are implemented in a central controller also called as the \textit{SDN Controller}. On the other hand, cognitive networking \cite{slavik_2008} refers to the network paradigm in which the networks automatically learn and respond to changes by actively taking decisions and planning network resources to achieve the end to end performance goals.
\khan{SDN offers the software adaptation to implement cognitive networks \cite{thomas_2006}}.
Cognitive networks can be realized using both distributed and centralized architectures. A novel approach to realize cognitive networks is to adapt data-driven machine learning (ML) algorithms to address challenges in future ultra-dense and dynamic networks \cite{zhang_2017, fadlullah_2017, ridha1}. ML algorithms can be used for both network design \cite{cadger_2017, del_2016, Abusubaih2022, Awad2021} and network performance evaluation \cite{khan_2016, kriara_2013, phit_2006, chinchali}. 
\par
This paper proposes a centralized network architecture using an SDN controller that uses machine learning algorithms to solve the two aforementioned network problems. Firstly, it anticipates the handover event that is likely to occur and to decide whether the handover is actually required? The proposed scheme reduces the likelihood of unnecessary handover \khan{events} in the presence of Overlapping BSS (OBSS) in ultra-dense deployment. Secondly, it solves the AP selection problem by predicting the post-selection network throughput to choose the best AP. \khan{Network} throughput is a significant metric to measure the user experience. The prior knowledge of future throughput can help to avoid network congestion \khan{which can play a vital role} in AP selection decision. The proposed scheme can be used to develop large frameworks and testbeds for real-time monitoring and network diagnostic to boost the \khan{Quality of Service (QoS)} in Wi-Fi networks.

\subsection{Contribution}
\label{subsec:contrib}
The paper presents efficient \khan{schemes} to address two fundamental issues (i.e., handover prediction and AP selection) in cognitive Wi-Fi networks. The proposed schemes uses data-driven machine learning algorithms to solve the two problems. The results are compared with traditional methods to validate the benefits of the proposed scheme. The paper also proposes useful datasets to train the machine learning algorithms for robust performance. The datasets used in this study are acquired using simulations performed in \khan{network simulator} (ns-3) \cite{ns3} and Mininet \cite{mininet} emulator. An interesting contribution is the use of \textit{ns-3 building} class in the simulation which is not used in previous related works to the best of our knowledge. The use of \textit{ns-3 building} class allows simulating both indoor and outdoor scenarios by configuring buildings, floors, rooms and other real-world structures in simulation to acquire more realistic datasets. \par

The rest of the paper is organized as follow: Section \ref{sec:rel_work} presents the state-of-the-art approaches to solve the handover prediction and AP selection problems. Section \ref{sec:prop_scheme}  describes the proposed scheme. Section \ref{sec:evaluation} explains the methodology used to evaluate the performance of the proposed scheme. The evaluation results are reported in Section \ref{sec:results}. \khan{Lastly}, conclusions are drawn in Section \ref{sec:conclusions}.

\section{Related Work}
\label{sec:rel_work}
\begin{table*}[htbp]
\centering
\caption{Summary of related works.}
\label{tab:rel_work}
\resizebox{\textwidth}{!}{

\input{tables/rel_work2.tex}
}
\end{table*}

A cognitive network is a modern network \khan{architecture} that is fully aware of the network state and can adapt to the varying network conditions. Such  a network \khan{learns} from these adaptations to make future decisions to achieve the end-to-end performance goals. Several prototypes for realizing cognitive networks are proposed in the last few years \cite{danieletto_2014, ali_2013, suresh_2012}. To realize efficient and scalable cognitive networks, machine learning techniques are being used in \cite{van_2009} and \cite{ayoubi_2018}. ML algorithms can be used for sophisticated learning and decisions-making in large and complex wireless networks where analytical methods do not meet the required performance \khan{requirements}. Machine learning techniques are applied to several problems in wireless networks, e.g. throughput estimation \cite{lin_2009, khan2020real}, interference classification \cite{kriara_2013}, delay analysis \cite{kajita_2014} and channel migration strategy \cite{kajita}. 
\par

In \cite{mariyam_2010}, authors address the problem of throughput estimation for TCP flows in Wide Area Networks (WAN). The authors used Support Vector Regression (SVR) on dataset obtained using a laboratory testbed. The prediction accuracy is evaluated using "relative prediction error" metric. In \cite{liu_2015}, authors \khan{address} the \khan{TCP throughput prediction} in cellular (3G/HSPA) networks using seven prediction algorithms and compared the prediction accuracy of each using the root-mean-squared error (RMSE) metric.
In \cite{samba_2016}, authors \khan{propose} a throughput estimation strategy in Long Term Evolution (LTE) cellular network using several network parameters such as RSSI, Signal-to-Noise Ratio (SNR), Reference Signal Received Quality (RSRQ) and Reference Signal Received Power (RSRP). The authors used three machine-learning algorithms namely Generalized Linear Model (GLM), Artificial Neural Networks (ANN) and Random Forests (RF), to evaluate the predictor performance. \par

Previous works on handover prediction in Wireless LANs \cite{park_2013, montavont_2006, mhatre_2006} use timeseries forecasting methods. Classical timeseries forecasting methods include autoregression (AR), moving average (MA), Autoregressive Moving Average (ARMA), Autoregressive Integrated Moving Average (ARIMA), Simple Exponential Smoothing (SES) and other variants of these methods. For instance, authors in \cite{park_2013} proposed a method to trigger handover using RSS based prediction in Wireless LANs. The authors \khan{argue} that the \khan{RSS value remains constant} during a short time interval (0.5 seconds in the proposed model), and predict the future RSS values using the autoregressive process of order 1 i.e. AR(1).
The proposed scheme is evaluated using dataset collected from ns-3 simulation. Handover event is predicted using the position information of the mobile device in \cite{montavont_2006}. In \cite{mhatre_2006}, authors proposed a handoff scheme based on the continuous monitoring of wireless links using the short-term and long-term trends in signal strength of beacon frames. The proposed scheme claims 50\% reduction in the handover delay as well as improvement in the overall performance. In \cite{kim_2007} authors used two RSS-based methods for handover prediction i.e. ARMA for stationary signals and ARIMA for non-stationary signals. 
\par

In \cite{yan_2008}, authors proposed traveling distance prediction based model for handover decision. Authors used the RSS values to calculate the distance between the AP and the mobile terminal (MT) using the following formula.
\begin{equation}
    RSS_P \; = \; E_t \times l_{OP}^{-\beta} \times 10^{\epsilon/10}
\end{equation}
where $E_t$, $\beta$ , $\epsilon$ represent the transmit power (in mW) of the AP, the path loss exponent and a zero mean Gaussian distributed random variable, respectively. The algorithm assumes that the MT travels at a constant speed. Results are compared with Mohanty's \cite{Mohanty2006} and Varma's methods \cite{Varma2003}.
The classical methods used in the aforementioned works, perform poorly on noisy data and in multi-step forecasting \cite{brownlee_2018}. Hence, there is an opportunity to use novel and efficient methods to solve the handover prediction problem. \khan{For instance, authors in \cite{Noura2019} proposed handover prediction using recurrent neural networks (RNN) in vehicular networks. Although RNN is relatively a more sophisticated method for time series predictions, we did not use RNN in our proposed scheme due to their complexity. Instead, we designed our dataset in a way to capture the time-dependency. Furthermore, the work \cite{Noura2019} does not considers AP load while selecting the new AP after handover.}.
\par

To solve the AP selection problem in dense networks, authors in \cite{oni_2017} propose a decentralized algorithm. The proposed Optimal AP Selection Algorithm (OPASA) uses the estimated downlink SINR which captures inter-BSS interference from overlapping APs. The authors show that OPASA outperforms \khan{the} SSF algorithm by achieving up to 99\% aggregate throughput gain.
In \cite{vasudevan_2005}, authors propose \textit{potential bandwidth} as a metric for AP selection. Potential bandwidth is calculated from the beacon timing of APs. Authors in \cite{bejerano_2004} solve the AP selection problem using approximation of Max-min fair bandwidth allocation algorithm. The authors compare the results and show that the proposed approximation algorithm outperforms both SSF and LLF algorithms. \khan{Authors in \cite{Militani2019} proposed to use supervised ML methods (i.e., naive bayes, decision trees, and random forest) to solve the AP selection, with higher throughput improvement reported for random forest method.}

\section{Proposed Scheme}
\label{sec:prop_scheme}

\subsection{Architecture}
\label{subsec:arch}
\begin{figure}[htbp]
\centering
\includegraphics[width=0.7\columnwidth]{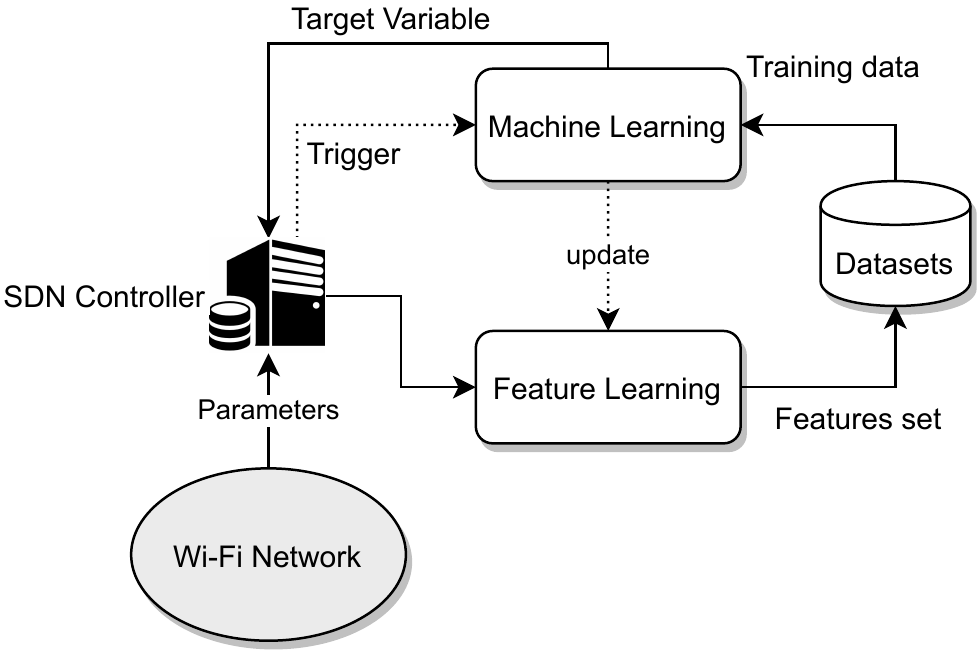}
\caption{Framework for cognitive network.} 
\label{fig:prop_scheme}
\end{figure}

The proposed scheme consists of four components: SDN controller, feature extraction module, datasets, and machine learning module. Fig. \ref{fig:prop_scheme} illustrates the functional architecture of the proposed scheme.\par

The SDN controller constantly collects network data consisting of several parameters of interest such as device's capability, supporting rates, battery status, \khan{user's} position and speed information, Wi-Fi channel being used, packet arrival rates, average throughput and frames retransmission ratios. The network attributes constitute raw data which is then processed to extract useful features. In the feature extraction module, some attributes are directly used as features, whereas some new features are created from the raw data. For instance, the number of associated clients to an AP is directly used as a feature, whereas the inter-arrival time of the packets is a feature that is computed from the packet-arrival times of two consecutive packets. \par

The features are then combined to form ML-ready datasets which are used by ML algorithms to implement end-to-end learning. Two types of datasets are created namely \textit{design datasets} and \textit{evaluation datasets}. The design datasets are used to predict a design parameter e.g. the AP for association. Other examples include the maximum number of nodes served by the AP, transmit power of the AP and the optimum channel to be used etc. The evaluation datasets are used for evaluating the network performance in the current conditions e.g. transmission throughput. Other examples include average packet end to end delay, packet inter-arrival rates, network congestion and channel access delay.
\par

\subsection{Functional Overview}
\label{subsec:functional_overview}
The SDN controller continuously \khan{monitors} the network changes (called as triggers). Three types of triggers are used by the controller i.e. (i) topology change, (ii) performance degradation, and (iii) periodic triggers. A new user sending association request to an AP corresponds to the first type of trigger. The lower network throughput or increase in the packet end-to-end delay than a pre-defined threshold \khan{level corresponds} to the second type of trigger. Periodic triggers are activated at regular intervals regardless of any change in the network state. The activation of any of these triggers automatically \khan{runs} the appropriate ML model. The ML model at fixed intervals imports the required ML-ready dataset form the database to retrain. When triggered, the ML model can thus generate accurate output. The output of the ML model is used by the SDN controller to implement a control action. The proposed scheme for handover prediction and AP selection is explained as follow:

\subsection{Handover Prediction Scheme}
\label{subsec:handover_pred_scheme}
\begin{figure}[htbp]
\resizebox{\columnwidth}{!}{
\centering
\input{figures/forecast.tex}
}
\caption{Handover prediction framework.} 
\label{fig:forecast}
\end{figure}
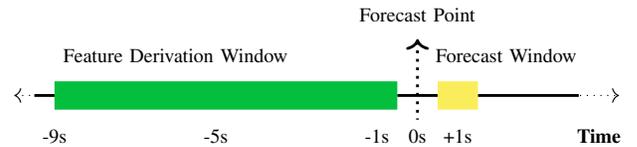

Handover prediction is solved as a binary classification problem using supervised learning techniques. The raw data for handover prediction consists of timeseries of RSS values of beacon frames received from APs. To be used in supervised learning, the timeseries is transformed into dataset as illustrated in Fig. \ref{fig:forecast} and further explained in this section.


\begin{figure}[htbp]
\centering
\includegraphics[width=0.6\columnwidth]{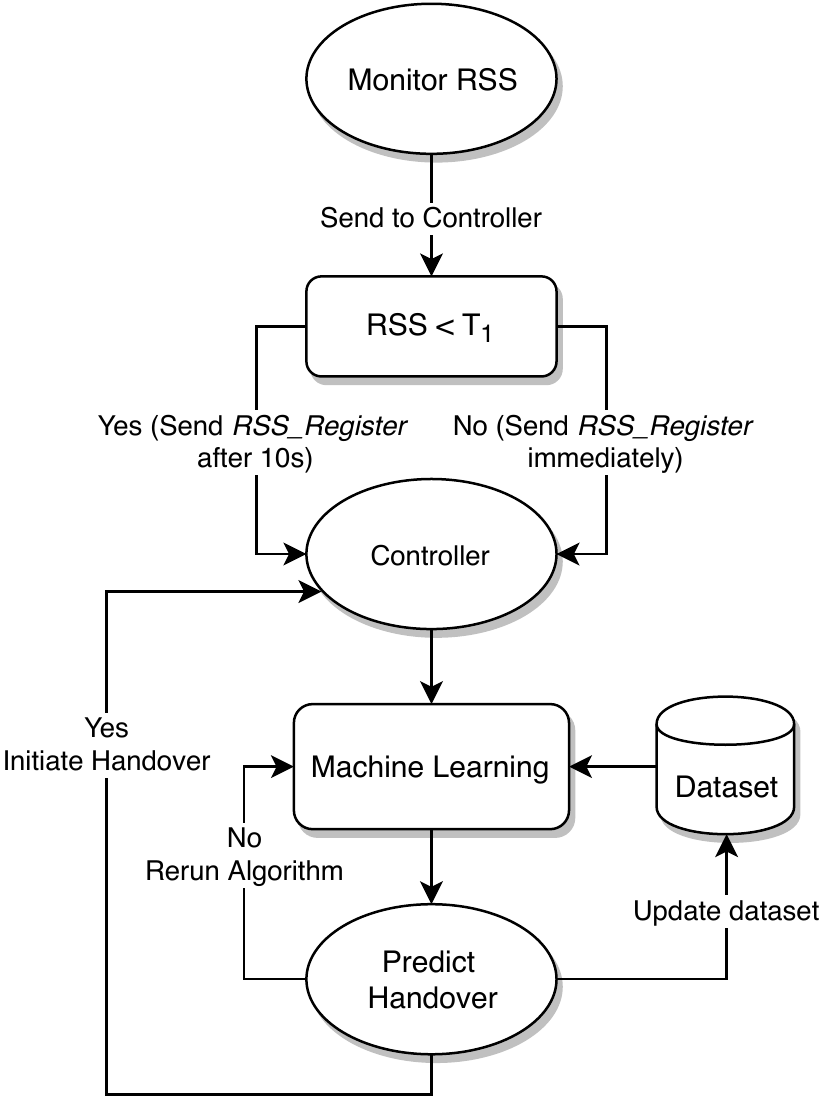}
\caption{Handover prediction scheme.} 
\label{fig:handover_pred_scheme}
\end{figure}

Fig. \ref{fig:handover_pred_scheme} illustrates the proposed handover prediction scheme. Each device constantly monitors the RSS and records the RSS values in beacon frames in a \textit{RSS\_REGISTER}. The \textit{RSS\_REGISTER} is then shared with the controller every second (beacon frequency). The controller copies the values from the \textit{RSS\_REGISTER} into a database of raw data. Each time a \textit{RSS\_REGISTER} is received, it is appended to the previous data. The raw data is then accessed by the feature extraction module, which transforms the raw data into ML-ready dataset. The \khan{ML-ready} dataset consists of several features as depicted in Table \ref{tab:dataset_ho}. Each sample in the dataset consists of 13 features. Columns 1 to 10 contains per-second average RSS values for 10 seconds. Column 11-13 contain the statistics calculated based on the first 10 columns i.e. mean, minimum and maximum. Each row in the dataset is calculated by applying a unit (1 second) shift to the previous column. 
\par
The controller constantly monitors the current association of the device. The method defines two RSS thresholds denoted as $T_1$ and $T_2$. $T_1$ refers to the RSS level which is significantly low, but still supports an ongoing connection despite if RSS \khan{drops} below it. Whereas, $T_2$ refers to the RSS level which is the minimum level to support a connection. If RSS drops slightly below the threshold, the connection will be terminated. The controller sends first trigger when the RSS of the device drops the first threshold $T_1$. The first trigger indicates the possibility of a handover in the next couple of seconds and hence a proactive measure is necessary. The trigger activates the machine learning module to run the algorithm at each time step to predict the probability of handover in the next time step. It is worthy to note that the first trigger is significant to reduce unnecessary processing by continuously running the ML algorithms when the device lies in good coverage. Once the trigger is generated, the ML module \khan{runs} the trained model to predict whether handover should \khan{be initiated or not}? The ML module periodically imports the most recent feature vector from the dataset for inference and \khan{runs} the model to predict the handover. The dataset is updated by appending the prediction decision for the given feature vector to improve the future learning process and prediction accuracy. \par

When the handover is detected for a given feature vector, the handover process is initiated. After completing the handover, when the RSS from the new AP is increased and becomes higher than $T_1$, the controller sends another trigger to the machine learning module to stop running the prediction process. If at anytime, the RSS drops to the second threshold $T_2$, a handover is initiated without running the ML model (the manual handover decision is not illustrated in Fig. \ref{fig:handover_pred_scheme}) and the dataset is updated by appending the handover decision to the given feature vector.

\begin{table}[h!]
\centering
\caption{Dataset for handover prediction.}
\label{tab:dataset_ho}

\input{tables/dataset_ho.tex}
\end{table}

\subsection{Access Point Selection Scheme}
\label{subsec:ap_select_scheme}
\begin{figure}[h!]
\centering
\includegraphics[width=0.7\columnwidth]{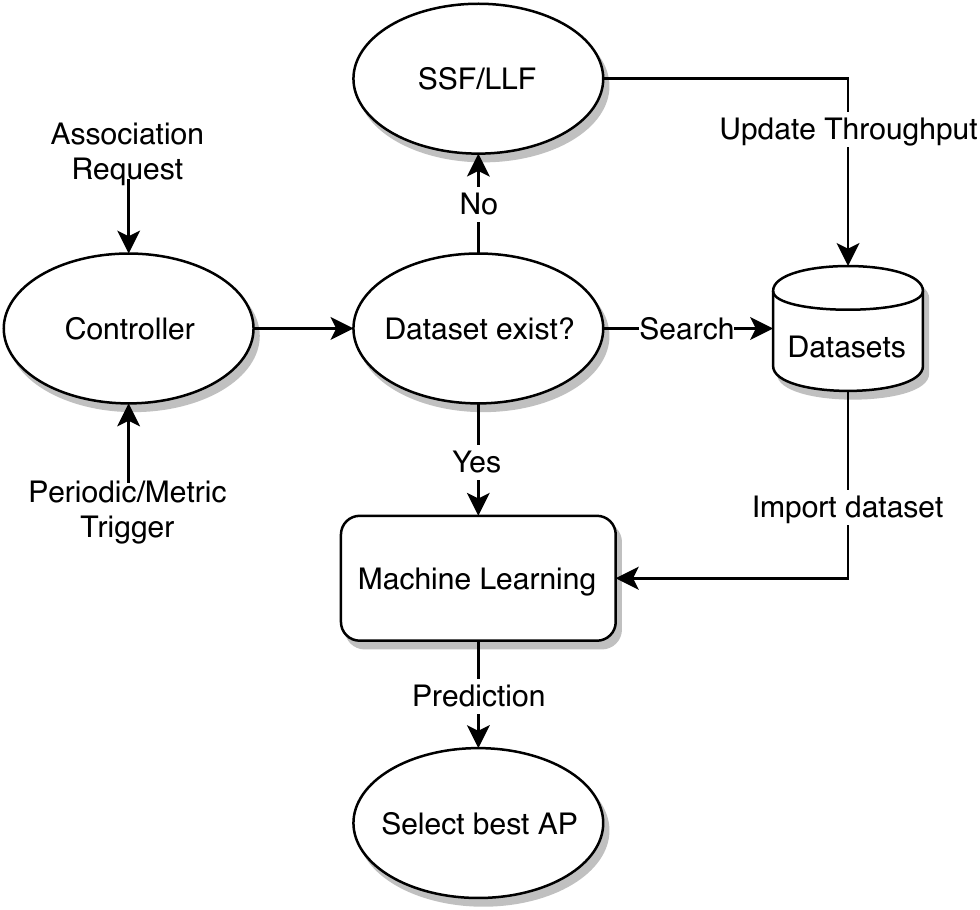}
\caption{AP selection flow diagram.}
\label{fig:ap_select_scheme}
\end{figure}

The AP selection problem is addressed by the proposed scheme using a multi-criteria online learning technique as illustrated in Fig. \ref{fig:ap_select_scheme}

When an AP receives an association request from a Wi-Fi station (STA), it forwards this request to the SDN controller. The SDN controller checks if the dataset (Table \ref{tab:dataset_ap}) is available to use machine learning algorithm to choose the best AP to offer connection to the new user. Initially when the network is first deployed, the dataset is not available. Hence, the controller uses the default algorithm (i.e. SSF or LLF) to select the AP. The controller computes the per BSS throughput for the given network parameters. Once, the dataset is populated with sufficient datapoints, any new association request is handled by the machine learning model. The proposed scheme predicts the throughput for each AP in the overlapping BSS and returns the estimated throughput for each AP (if the new STA would be associated to this AP) to the controller. The controller then selects the AP which provides higher estimated throughput for connecting the requesting client.
\par
To create the dataset for throughput estimation, the controller constantly records the information such as the number of associated clients and packet information (e.g. timestamps, arrival time, packet size and signal to noise ratio etc.). New feature, Inter-Arrival Time (IAT) is calculated from the timestamp and arrival time of each packet. The two features, IAT and the number of clients connected to the access point are primarily selected to be used for throughput estimation. Furthermore, new features are derived from the IAT values, using the statistics such as \khan{minimum, maximum, mean, variance, skew and kurtosis}. The features are collected over a time window of fixed duration for the whole network. The structure of ML-ready dataset for throughput estimation is given in Table \ref{tab:dataset_th}
\begin{table*}[!h]
\centering
\footnotesize
\caption{Dataset for throughput estimation.}
\renewcommand{\arraystretch}{0.7}
\label{tab:dataset_th}

\input{tables/dataset_th.tex}

\end{table*}

For AP selection, the controller simultaneously collects other parameters to compute features to create dataset. The structure of dataset used for AP selection is listed in Table \ref{tab:dataset_ap}.
\par

\begin{table*}[!h]
\centering
\footnotesize
\renewcommand{\arraystretch}{0.7}
\caption{Dataset for AP selection.}
\label{tab:dataset_ap}

\input{tables/dataset_ap.tex}

\end{table*}

\section{Evaluations}
\label{sec:evaluation}
The proposed scheme is implemented using ns-3 simulator \cite{ns3} and Linux-based Mininet network emulator \cite{mininet}. Mininet provides a sufficient level of flexibility and control over the network to dynamically implement new configurations. Additionally, it allows interactive simulation and user can add traffic and applications on devices as well as apply some topological changes during the simulation runtime, thus enabling users to create more dynamic scenarios. On the other hand, ns-3 is a de-facto standard for simulating wireless networks. It provides accurate models of the wireless channel. The recent version of ns-3 also supports indoor models where users can model buildings, floors, rooms and other parameters of the real world.
\par
To implement the proposed scheme for handover prediction, we performed extensive simulations in ns-3 to acquire raw network data. Both indoor and outdoors devices are deployed in the simulation. The simulation uses the design parameters defined in Table \ref{tab:sim_para}.

\begin{table}[h]
\centering
\renewcommand{\arraystretch}{0.9}
\caption{Simulation parameters for building topology.}
\label{tab:sim_para}
\input{tables/sim_para.tex}
\end{table}

The raw data acquired is transformed into dataset as given in Table \ref{tab:dataset_ho}. The datasets are then used in Mininet-based simulation to predict handovers using Random Forest (RF) algorithm. Random Forest (RF) \cite{liaw_2002} is a supervised learning algorithm employed in classification problems. It randomly selects features to build several decision trees and then averages the results. It is relatively a simpler algorithm and requires less time to train a ML model.

To implement the proposed scheme for AP selection, the controller is configured to simulate the two user association algorithms i.e. SSF and LLF in Mininet. The simulations include 3 APs and 50 STAs, randomly moving in the network and changing association controlled by these algorithm. The network traces are collected and dataset is created according to Table \ref{tab:dataset_ap}. 
Lastly, the controller \khan{uses an ML} model to perform AP selection. The previously collected datasets are used to train the ML model to estimate network throughput. The STA-AP association which will give higher estimated aggregate throughput, is then selected.

The AP selection dataset involves the use of estimated throughput and hence it is necessary to evaluate the accuracy of the algorithms which estimates the throughput. To evaluate the accuracy of estimated throughput, we used two algorithms i.e. \khan{Multi-layer Perceptron} (MLP) and SVR due to their capability to better predict such metrics  \cite{ayoubi_2018}. The raw traces form the simulated network are collected and transformed into useful features as listed in Table \ref{tab:dataset_th} to create the ML-ready dataset. The dataset is divided into training-validation (70-30 \%) splits. The two algorithms are trained with the training data and are then tested by applying to the unseen validation data. To further validate the statistical significance of the model, 10-fold cross validation is used to avoid over-fitting. 
\par

\section{Results and Discussion}
\label{sec:results}

The performance of the proposed handover prediction scheme primarily depends on the accuracy of the machine learning model. Firstly, the prediction accuracy of the RF algorithm used for handover prediction is evaluated using confusion matrix. A confusion matrix is used to evaluate the percentage of correct and wrong predictions on data points of all classes in the dataset. The confusion matrix shown in Table \ref{tab:confusion_matrix} shows the accuracy of the RF algorithm.

\begin{table}[!h]
\centering
\caption{Confusion matrix.}
\label{tab:confusion_matrix}
\input{tables/confusion_matrix.tex}
\end{table}

It can be seen that the RF algorithm provides high accuracy to correctly predict the handover events. In the next step, the performance of the proposed handover prediction scheme is compared to other methods stated earlier to assess the overall performance.
Fig. [7] shows the performance of the proposed scheme versus two other handover prediction methods based on RSS forecasting method \cite{park_2013} and traveling distance method \cite{yan_2008}. The figure shows the number of unnecessary handovers (cumulative) over time, computed for the three methods. It can be seen that the proposed scheme outperforms the two methods by reducing the overall numbers of unnecessary handover. The analysis of results show that the proposed scheme reduces the number of unnecessary handovers by approximately 60\% and 50\% as compared to RSS method and traveling distance method, respectively.

\begin{figure}[!h]
\centering
\includegraphics[width=0.7\columnwidth]{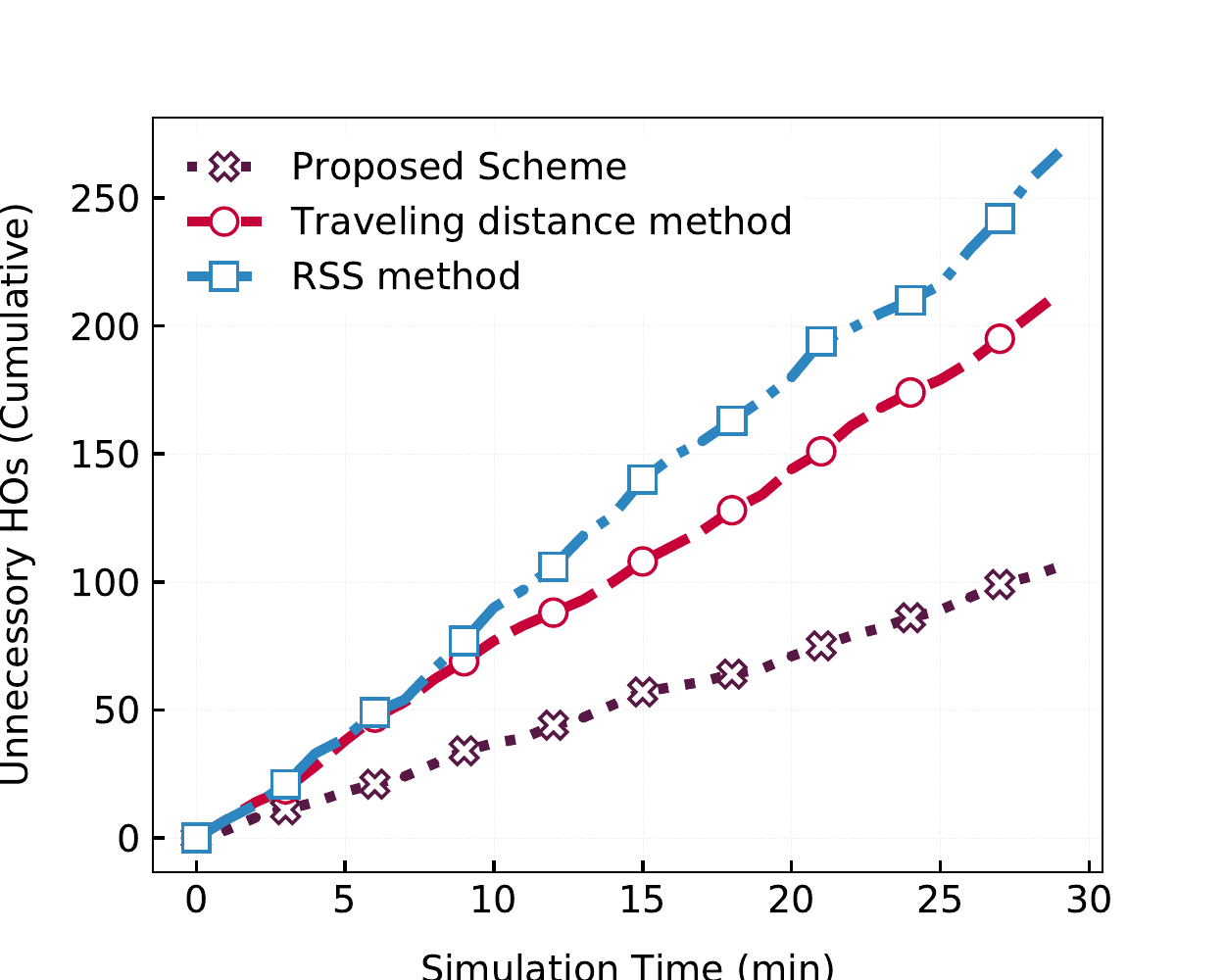}
\caption{Unnecessary handovers using the proposed scheme versus previous approaches.}
\label{fig:handover_result1}
\end{figure}

The proposed scheme for AP selection problem is then evaluated. As the accuracy of the AP selection scheme directly relies upon the accuracy of the throughput estimation, we first evaluate the accuracy of the throughput estimation using two ML algorithms i.e. MLP and SVR. 

\begin{figure}[!h] 
\centering
\includegraphics[width=0.7\columnwidth]{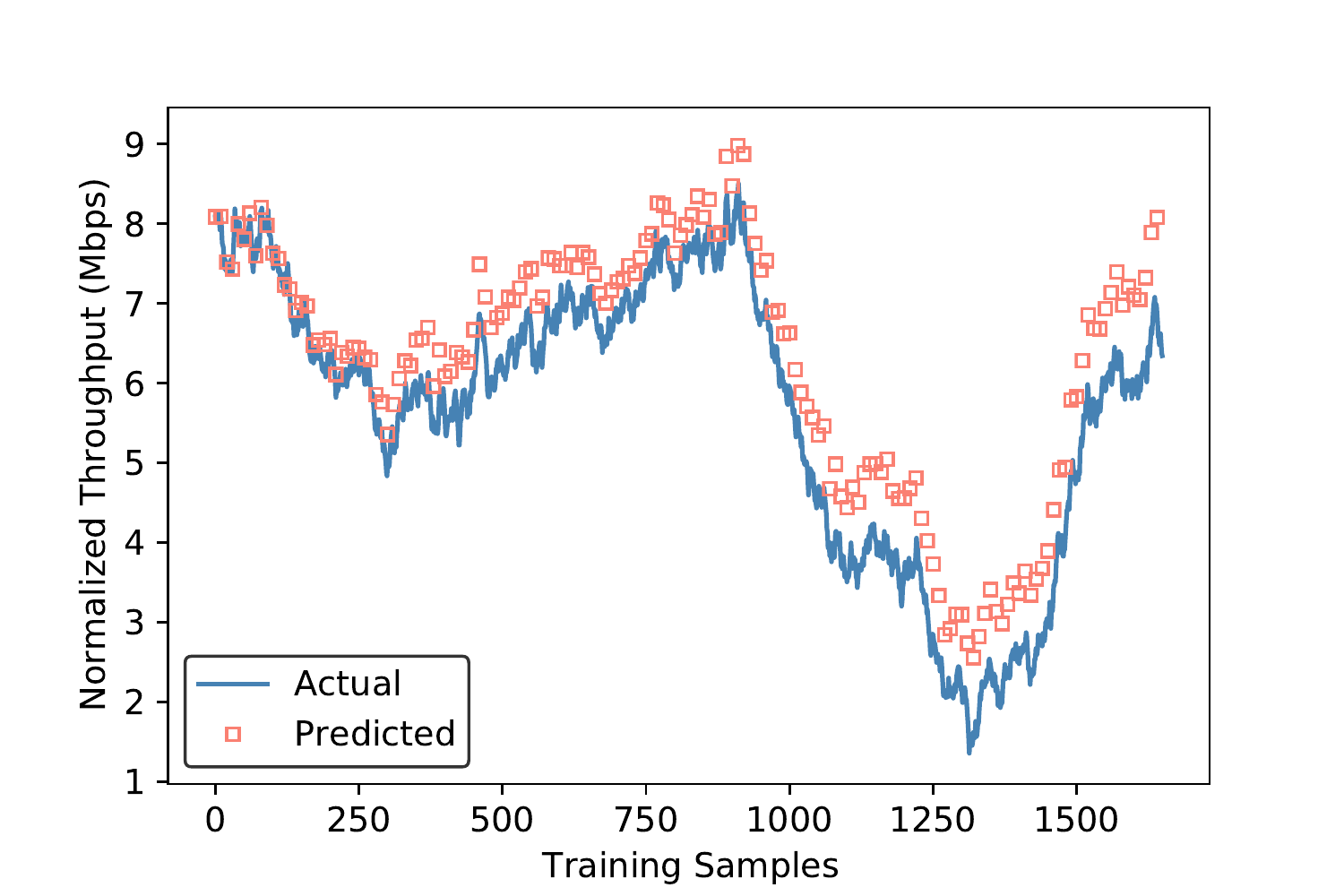}
\caption{Throughput estimation (using MLP).}
\label{fig:throughput_mlp}
\end{figure}

\begin{figure}[!h] 
\centering
\includegraphics[width=0.7\columnwidth]{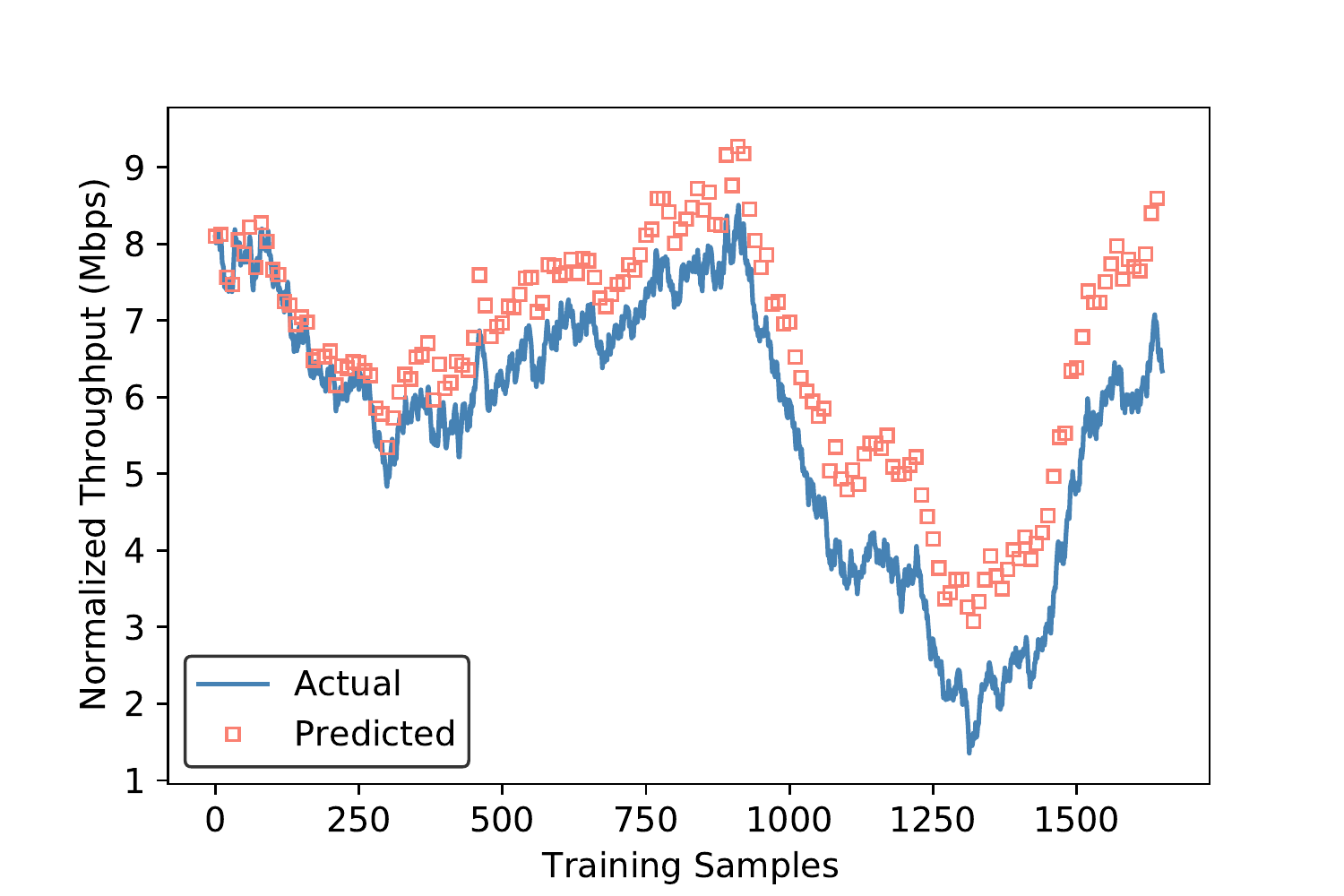}
\caption{Throughput estimation (using SVR).}
\label{fig:throughput_svr}
\end{figure}

The predicted throughput versus actual throughput is plotted for both algorithms as given in Fig. \ref{fig:throughput_mlp} and \ref{fig:throughput_svr}. It can be observed that the MLP model provides better accuracy (i.e. predicted values are much closer to the actual values) as compared to the SVR model.
To further quantify the performance of both models, three performance metrics i.e. training time, Mean Squared Error (MSE) and R-squared are computed and the results are listed in Table \ref{tab:throughput_comp}. The MLP based model requires long training time (1.59 second) than the SVR model (0.211 seconds), however it provides better accuracy (i.e. less MSE for MLP = 0.067 as compared to SVR = 0.211) and better generalization to future predictions (i.e. higher R-squared for MLP = 0.974 as compared to SVR = 0.916). The better learning capabilities of MLP costs longer training time due to its complex design (hundreds of neurons arranged in several layers).

\begin{table}[!h] 
\centering
\small
\caption{Performance and complexity analysis of throughput estimation models.}
\label{tab:throughput_comp}
\renewcommand{\arraystretch}{0.7}
\input{tables/results_throughput.tex}

\end{table}

\begin{figure}
\centering
\subcaptionbox{Average throughput.}{\includegraphics[width=0.45\columnwidth]{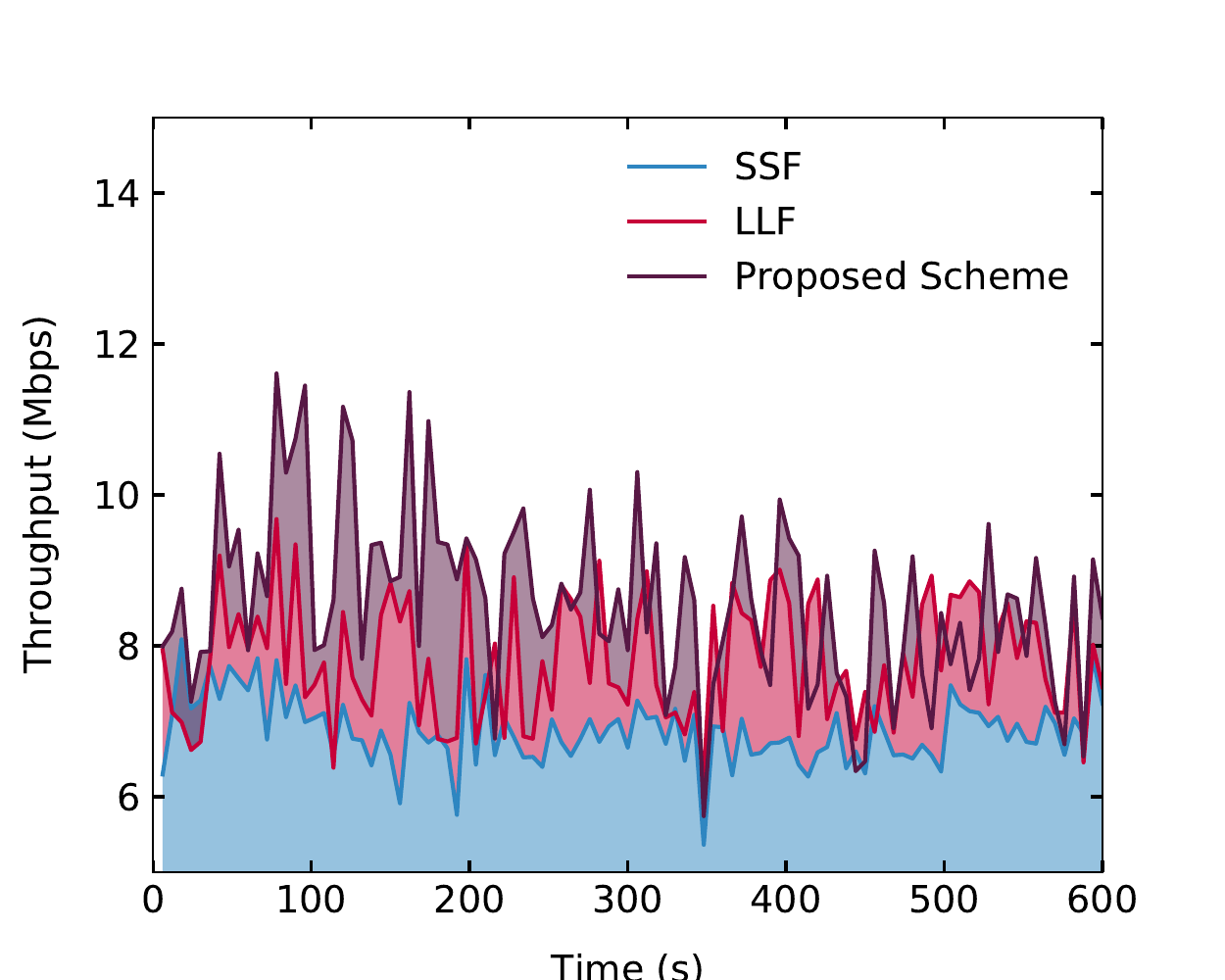}}
\subcaptionbox{Cumulative throughput.}{\includegraphics[width=0.45\columnwidth]{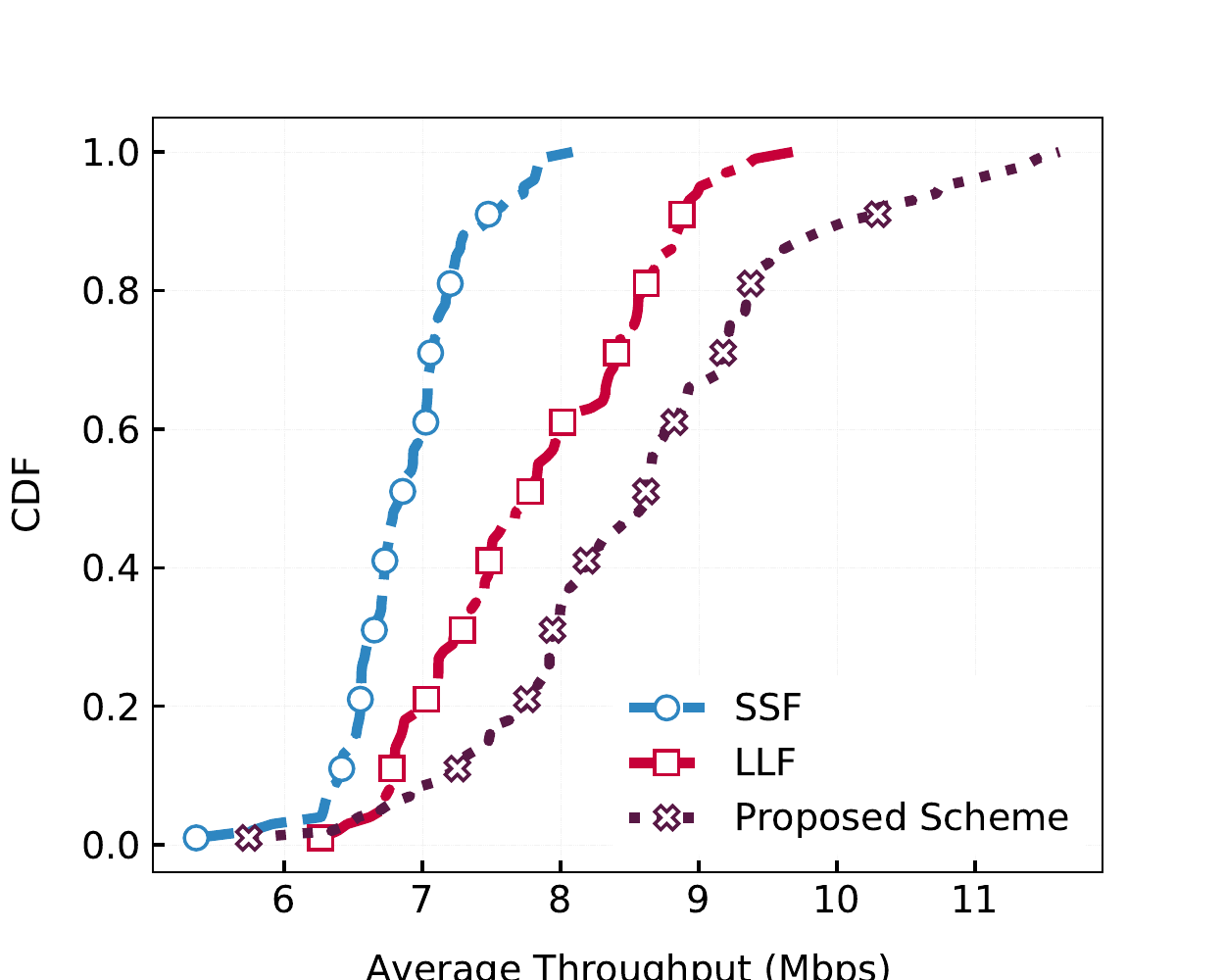}}
\caption{Comparison of the average BSS throughput.}
\label{fig:average_bss_throughput}
\end{figure}

\begin{figure}[!h]
\centering
\includegraphics[width=0.7\columnwidth]{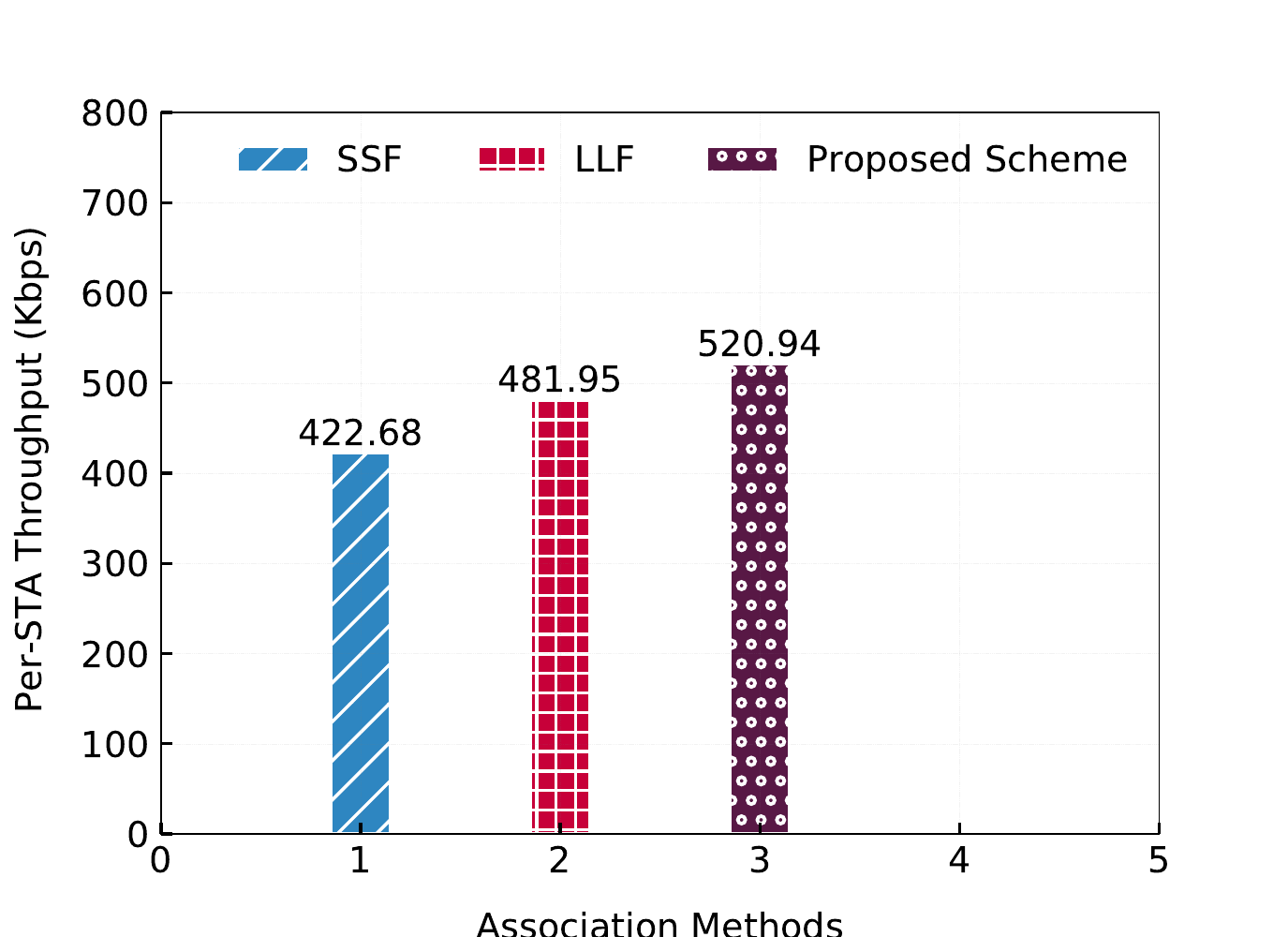}
\caption{Comparison of per-STA throughput.}
\label{fig:per_sta_throughput}
\end{figure}

The estimated throughput using MLP algorithm is then used for AP selection. In AP selection, two performance metrics i.e. average BSS throughput and per-STA throughput are used to compare the throughput gain of the proposed scheme versus \khan{standard} AP selection schemes (i.e. SSF and LLF). The results are shown in Fig. \ref{fig:average_bss_throughput} (average BSS throughput) and \ref{fig:per_sta_throughput} (per-STA throughput). It can be observed that the proposed scheme improves the average BSS throughput as well as per-STA throughput. The analysis of throughput gains report an average improvement of 9.2\% and 8\% as compared to the SSF and LLF schemes respectively. \khan{It is worthy to note here, that the work in \cite{bejerano_2004} also propose an alternate AP association scheme for load balancing in Wi-Fi networks and compared against SSF and LLF schemes. We could not compare our work against \cite{bejerano_2004} due to the complexity of the scheme in ns-3 environment. Additionally, it would not be fair to compare our results against \cite{bejerano_2004} due to the difference in network configurations, topology, and the assumptions typically involved in analytical models versus simulation environments.}

\section{Conclusions and Future Work}
\label{sec:conclusions}
The paper proposes machine learning techniques to solve two well-known problems in WLAN networks i.e. the handover prediction problem and the AP selection problem. The handover prediction problem is \khan{formulated as} a multi-step time-series prediction problem. It is solved using supervised learning algorithm i.e. random forest in the proposed scheme. The goal in binary prediction problem is to achieve high prediction accuracy. On the contrary, the AP selection is a design problem to find the optimum AP-STA associations that improves the network throughput performance. The proposed scheme solves this issue by estimating throughput using all possible configurations and selects the one which provides higher throughput gain. The performance of the proposed scheme is validated and results report significant improvement in the overall performance. The proposed scheme for handover prediction outperforms the RSS method and traveling distance method by reducing the number of unnecessary handovers by 60\% and 50\% respectively. In the AP selection problem, the proposed scheme outperforms the SSF and LLF algorithms by achieving higher throughput gains upto 9.2\% and 8\% respectively. However, it is expected that a large set of problems and challenges in future Wi-Fi networks can be solved using similar approaches. Although running ML applications over resource constrained mobile devices can be challenging, the new edge computing paradigm \cite{ishtiaq2021edge} can be a promising approach to meet the computation requirements of future networks.

\bibliography{manuscript-R1.bib}
\bibliographystyle{ieeetr}

\vskip 0pt plus -1fil

\input{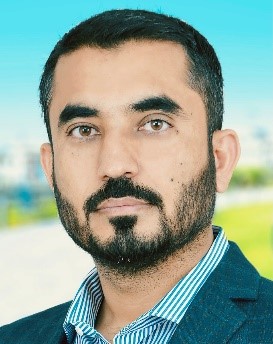}
\vskip 0pt plus -1fil

\input{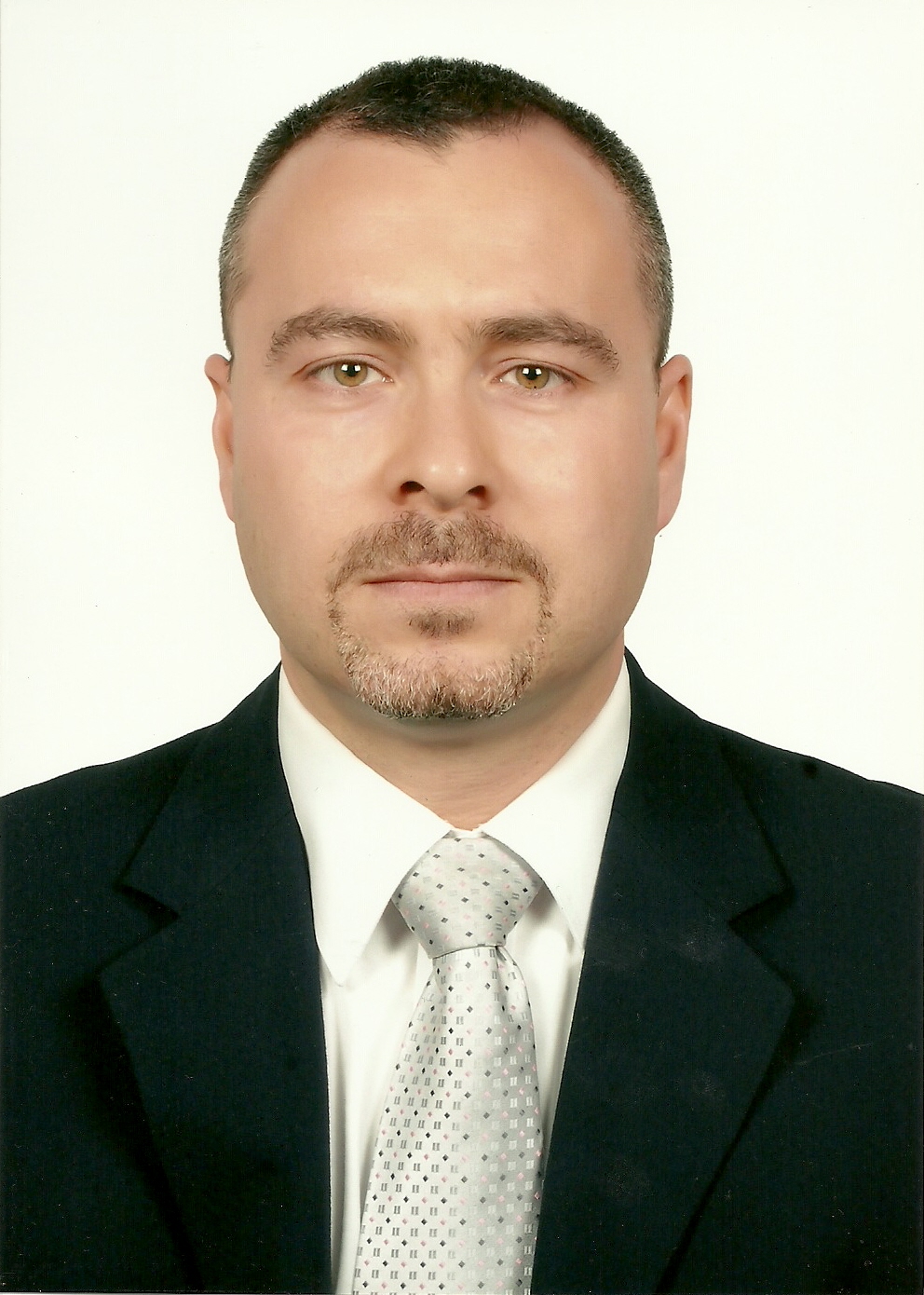}
\vskip 0pt plus -1fil

\input{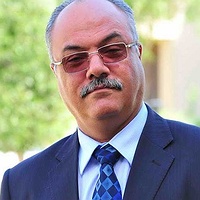}
\vskip 0pt plus -1fil

\input{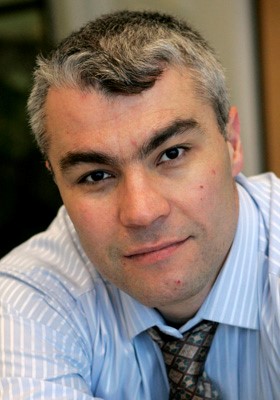}
\vskip 0pt plus -1fil

\input{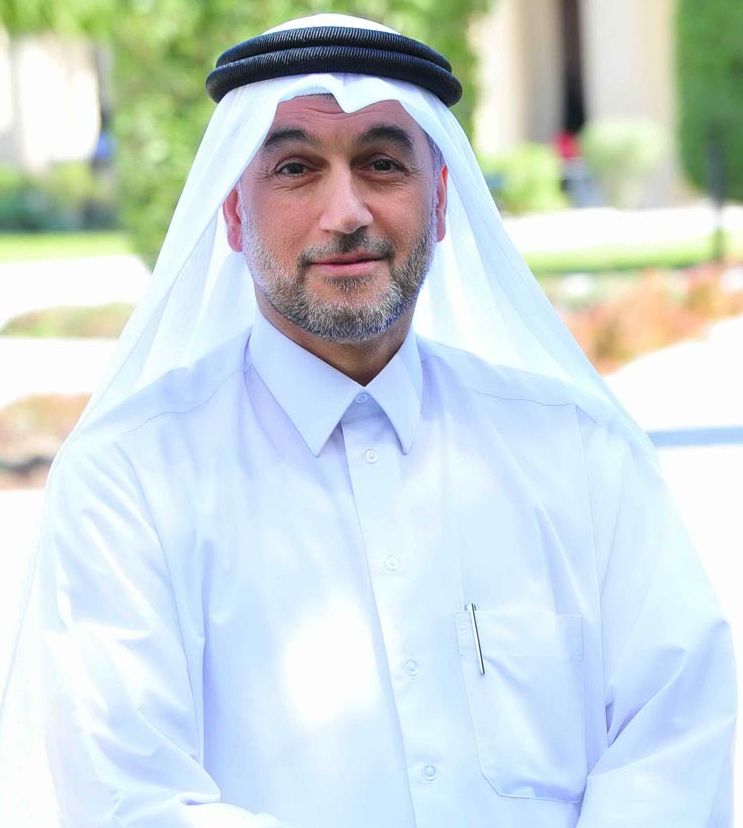}
\balance

\end{document}

%% file: figures/style.tex
\usepackage{tikz}
\usetikzlibrary{shapes.geometric, 
                arrows, 
                fit, 
                shapes, 
                calc, 
                patterns,
                positioning, 
                automata,
                decorations.pathreplacing, 
                backgrounds,
                shapes.misc,
                shapes.multipart,
                chains,
                arrows.meta,
                shadows,
                snakes,
                bending}
\usepackage{pgflibraryarrows} 

\definecolor{color1}{rgb}{0.64, 0.68, 0.82}
\definecolor{color2}{rgb}{0.52, 0.52, 0.51}
\definecolor{color3}{rgb}{1.0, 0.97, 0.86}
\definecolor{color4}{rgb}{1.0, 0.77, 0.05}
\definecolor{color5}{rgb}{0.55, 0.52, 0.54}
\definecolor{white}{rgb}{1.0, 1.0, 1.0}
\definecolor{color6}{rgb}{0.63, 0.79, 0.95}  
\definecolor{line}{rgb}{0.4, 0.26, 0.13}
\definecolor{color7}{rgb}{0.85, 0.75, 0.85} 
\definecolor{row}{rgb}{0.4, 0.6, 0.8}
\definecolor{white}{rgb}{1, 1, 1}

\pgfdeclarelayer{background}
\pgfsetlayers{background,main}
\pgfdeclarelayer{foreground}

\tikzstyle{startstop} = [rectangle, 
rounded corners, 
minimum width=2.5cm, 
minimum height=1.2cm,
align=center,
text centered, 
draw=black, 
text width=5em, 
line width=1pt,
fill=color6,  
drop shadow
] 

\tikzstyle{io} = [trapezium, 
trapezium left angle=80, 
trapezium right angle=100, 
minimum width=1cm,
minimum height=1cm, 
text centered, 
draw=black,
text width=1cm, 
line width=1pt,
fill=color6,
drop shadow
] 

\tikzstyle{process} = [rectangle, 
rounded corners, 
minimum width=2.5cm, 
minimum height=1.2cm, 
text centered, 
draw=black,
line width=1pt,
fill=white,
drop shadow] 

\tikzstyle{nw} = [ellipse, 
rounded corners, 
minimum size=2cm, 
text centered, 
draw=black,
text width=2cm, 
line width=1pt,
fill=white,
drop shadow] 

\tikzstyle{state} = [ellipse, 
rounded corners, 
minimum size=1cm, 
text centered, 
draw=black,
text width=1.5cm, 
line width=1pt,
fill=white,
drop shadow] 

\tikzstyle{decision} = [diamond, 
minimum width=2.5cm, 
minimum height=1cm, 
text centered, 
draw=black,
text width=5em, 
line width=1pt,                             fill=white, 
drop shadow
] 

\tikzstyle{text} = [rectangle, minimum width=3cm, minimum height=1cm, text centered, draw=black] 

\tikzstyle{arrow} = [thick,->,>=stealth]{}

\tikzstyle{wave}=[decorate, decoration={expanding waves, angle=7,
                          segment length=2}]

\tikzstyle{queue}=[
  rectangle split,
  minimum height=1cm,
  rectangle split horizontal,
  rectangle split parts=6, 
  draw, 
  anchor=center,
  line width=1pt,
  fill=white, 
  drop shadow
  ]{}

\tikzstyle{database}=[
      cylinder,
      shape border rotate=90,
      minimum height=1.5cm,
      minimum width=1.5cm,
      text width=1.5cm,
      aspect=0.25,
      draw,
      line width=1pt,
  	  fill=white, 
  	  drop shadow
     ]{}

\tikzstyle{ground}=[
		ellipse,
        minimum height=2cm,
        minimum width=6.0cm,
    	align=center]{}

\tikzstyle{round}=[
		circle,
        draw,
        minimum size=2cm,
    	align=center]{}
        
\tikzstyle{arrow1}=[line width=1mm,draw=black,-triangle 45,postaction={draw, line width=3mm, shorten >=4mm, -}]


%% file: tables/rel_work2.tex

\begin{tabular}{lllp{4cm}lll}

\toprule

Ref
& Category
& ML-based
& Parameters
& Algorithms
& Complexity
& Applications
\\
\midrule

\cite{samba_2016}
& Throughput Prediction
& Yes
& location, speed, SNR, RSRQ, RSRP, and RSSI
& GLM, NN, RF
& High
& File transfer
\\[1em]

\cite{liu_2015}  
& Throughput Prediction 
& Yes
& Past Throughput values
& Regression, NN, \newline SVR, EWMA
& Low
& TCP Session throughput
\\[1em]

\cite{lin_2009}
& Throughput Prediction
& Yes
& Frame Size and Channel SNR Values
& MLP
& Medium
& Frame Size Optimization
\\[1em]

\cite{kriara_2013}
& Throughput Prediction
& Yes
& MCS Index, Spatial Streams, Modulation Scheme, Coding Rate, Data Rates
& Categorical \newline Regression
& High
& Interference Classification
\\[1em]

\cite{kajita_2014}
& Throughput Prediction
& Yes
& Normalized RSS, Traffic Bytes transmitted by interfering stations, Inter-Channel Distance
& Linear \newline Regression
& High
& Channel selection
\\[1em]

\cite{park_2013}
& Handover Prediction
& No
& RSS
& AR(1)
& Low
& Inter-BSS handover
\\[1em]

\cite{montavont_2006}
& Handover Prediction
& No
& RSS, Position
& Monitoring RSS trends
& Low
& Inter-BSS handover
\\[1em]

\cite{kim_2007}
& Handover Prediction
& No
& RSS
& ARMA, ARIMA
& Low
& Inter-BSS handover
\\[1em]

\cite{yan_2008}
& Handover Prediction
& No
& RSS, distance
& Rate of RSS change
& Medium
& WLAN to cellular
\\[1em]

\cite{oni_2017}
& AP Selection
& No
& SINR
& OPASA
& High
& AP Selection in Dense WLANs
\\[1em]

\cite{vasudevan_2005}
& AP Selection
& No
& potential bandwidth
& Analytical approximation
& Low
& AP Selection in WLANs
\\[1em]

\cite{bejerano_2004}
& AP Selection
& No
& bandwidth
& Max-min Approximation
& High
& AP Selection in WLANs
\\[1em]

\bottomrule
\rowcolor{white}
\multicolumn{7}{l}    {}\\[0.01em]
\rowcolor{white}
\multicolumn{4}{l}    {\textbf{Terms:}}\\[0.4em] \rowcolor{white}
\multicolumn{4}{l}    {\textbf{RSSI:} Received Signal Strength Indicator}
&\multicolumn{3}{l}   {\textbf{RSRP:} Reference Signal Received Power} \\\rowcolor{white}
\multicolumn{4}{l}    {\textbf{RSRQ:} Reference Signal Received Quality}
&\multicolumn{3}{l}   {\textbf{GLM:}  Generalized Linear Models} \\\rowcolor{white}
\multicolumn{4}{l}    {\textbf{NN:}   Neural Network}
&\multicolumn{3}{l}   {\textbf{EWMA:} Exponential Weighted Moving Average} \\\rowcolor{white}
\multicolumn{4}{l}    {\textbf{SVR:}  Support Vector Regression}
&\multicolumn{3}{l}   {\textbf{RF:}   Random Forest} \\\rowcolor{white}
\multicolumn{4}{l}    {\textbf{MLP:}  Multi-Layer Perceptron}
&\multicolumn{3}{l}   {\textbf{AR:}   Auto Regression} \\\rowcolor{white}
\multicolumn{4}{l}    {\textbf{ARMA:} Auto-Regressive Moving Average}
&\multicolumn{3}{l}   {\textbf{ARIMA:} Auto-Regressive Integrated Moving Average} \\\rowcolor{white}

\bottomrule
\end{tabular}

%% file: figures/forecast.tex
\definecolor{yellow}{rgb}{0.98, 0.93, 0.36}
\definecolor{green}{rgb}{0.01, 0.75, 0.24}
\begin{tikzpicture}[]
\draw  [dotted, <->] (-5,0) -- (2.5,0);
\draw  [solid, line width=1] (-4.75,0) -- (-4.5,0);
\draw  [solid, line width=1] (0.75,0) -- (2,0);
\draw  [solid, line width=1] (-0.25,0) -- (0.25,0);
\draw  [solid, line width=1] (0.75,0) -- (0.5,0);

\draw [solid, color=green][line width=10] (-4.5,0) -- (-0.25,0) ;
\draw [solid, color=yellow][line width=10] (0.25,0) -- (0.75,0) ;
\draw  [dotted, <-, line width=1] (0,0.7) -- (0,-0.3);

\draw (-3,0.5) node  [align=left] {\scriptsize Feature Derivation Window};
\draw (1.1,0.5) node  [align=left] {\scriptsize Forecast Window};
\draw (0,1.0) node  [align=left] {\scriptsize Forecast Point};
\draw (2.25,-0.5) node  [align=left] { \textbf{\scriptsize Time}};

\draw (-4.5, -0.5) node  [align=left] {\scriptsize -9s};
\draw (-2.5,-0.5) node  [align=left] {\scriptsize -5s};
\draw (-0.5,-0.5) node  [align=left] {\scriptsize -1s};
\draw (0.5,-0.5) node  [align=left] { \scriptsize +1s};
\draw (0,-0.5) node  [align=left] { \scriptsize 0s};

\end{tikzpicture}

%% file: tables/dataset_ho.tex
\begin{tabular}{llllp{2cm}} 
\hline
\multicolumn{4}{l}{Features} 
& \multicolumn{1}{l}{Target Variable*}  \\[0.5em]
\cmidrule{1-4} \cmidrule{5-5}

\rowcolor[HTML]{FFFFFF}
1, 2,... 10               
& 11  & 12  & 13  & Binary variable   \\[0.5em] 
\cmidrule(lr){1-4} \cmidrule(lr){5-5}

RSS0 ... RSS9      
& Min    & Max    & Mean  
& HO prediction* \newline (0, 1)
\\
\bottomrule
\end{tabular}



%% file: tables/dataset_th.tex
\begin{tabular}{lllll}
\toprule

&Parameters  
&Derived                    
&States                 
&Data Type                   
\\[0.5em]
\midrule

\multirow{3}{*}{Features}

& Clients      
& n\_clients                                   
& -                
& integer                
\\

&Timestamp    
& \multirow{2}{*}{IAT} 
& [mean, min, max, skew, kurtosis]            
& float                  
\\

&Arrival Time
&&&            
\\ [1em] \midrule

\multirow{2}{*}{Target Variable}

&Arrival Time & \multirow{2}{*}{Throughput*}                 
& \multirow{2}{*}{-} 
& \multirow{2}{*}{float} 
\\

&Packet Size 
& & &  
\\[1em]
\bottomrule
\end{tabular}


%% file: tables/dataset_ap.tex
\begin{tabular}{lllll}
\toprule

&Parameters  
&Derived                    
&States                 
&Data Type                   
\\[0.5em]
\midrule

\multirow{5}{*}{Features}

& Clients      
& n\_clients                                   
& -                
& integer                
\\

&RSSI
& \multirow{2}{*}{SNR} 
& [mean, min, max, skew, kurtosis]           
& float                  
\\

&Noise Level
&                                              
&                       
&                        
\\[0.5em]


&MAC Queue Length
& \multirow{2}{*}{MAC delay} 
& [mean, min, max, skew, kurtosis]           
& float                  
\\

&Time stamp
&                                              
&                       
&                        
\\[1em] \midrule


\multirow{2}{*}{Target Variable}


&Arrival Time & \multirow{2}{*}{Throughput}                 
& \multirow{2}{*}{-} 
& \multirow{2}{*}{float} 
\\

&Packet Size 
&  &  &  
\\
\bottomrule
\end{tabular}

%% file: tables/sim_para.tex
\begin{tabular}{l l}
\toprule
Parameter        & Value                  \\
\midrule
Building Area ($m^2$)     & 300 x 100                       \\
No. of floors             & 1                               \\
No. of rooms along x-axis & 30                              \\
No. of rooms along y-axis & 10                              \\
Type of building          & Commercial                      \\
Wall type                 & Concrete with windows           \\
Mobility Model            & MobilityBuildingInfo            \\
Propagation Loss Model    & OhBuildingsPropagationLossModel \\
External Wall Loss (EWL)  & 7 dB \\  
\bottomrule
\end{tabular}

%% file: tables/confusion_matrix.tex
\begin{tabular}{c c c c}
& 
& \multicolumn{2}{c}{\textbf{Predicted}} 
\\[1em] 
& \multicolumn{1}{c}{}            
& \multicolumn{1}{c}{Handover}                     
& \multicolumn{1}{c}{No Handover}                  
\\ [0.5em]

\multicolumn{1}{c}{}                                  
& \multicolumn{1}{c}{Handover}    
& \multicolumn{1}{c}{\cellcolor[HTML]{A6E178}90.5\%} 
& \multicolumn{1}{c}{\cellcolor[HTML]{fa8072}9.5\%}  
\\[1em]

\multicolumn{1}{c}{\multirow{-2}{*}{\textbf{Actual}}} 
& \multicolumn{1}{c}{No Handover} 
& \multicolumn{1}{c}{\cellcolor[HTML]{fa8072}6.2\%} 
& \multicolumn{1}{c}{\cellcolor[HTML]{A6E178}93.8\%} 
\\ [1em]
\end{tabular}

%% file: tables/results_throughput.tex
\begin{tabular}{lll}
\toprule
Parameter
& MLP
& SVM
\\[1em]
\midrule


Training time (s)
& 1.59
& 0.211
\\[1em]

R-Squared
& 0.974
& 0.916
\\[1em]

MSE
& 0.067
& 0.211
\\[1em]
\bottomrule
\end{tabular}

%% file: biographies/khan.tex
\begin{IEEEbiography}[{\includegraphics[width=1in,height=1.25in,clip,keepaspectratio]{khan.jpg}}]{Muhammad Asif Khan} (SM'20) is a Postdoctoral Research Fellow at Qatar Mobility Innovations Center (QMIC), Qatar University. 
He received the Ph.D. degree in Electrical Engineering from Qatar University (2020), MSc degree in Telecommunication Engineering from University of Engineering and Technology Taxila, Pakistan (2013), and B.Sc. degree in Telecommunication Engineering from University of Engineering and Technology Peshawar, Pakistan (2009). He also worked as a Researcher Assistant at Qatar University (2014-2015) and at QMIC (2016-2017). His current research interests include wireless networks, edge computing, distributed machine learning, and computer vision. He is a Senior Member of IEEE, and Member of IET. For more detailed information, please visit his homepage: \url{https://asifk.me}.
\end{IEEEbiography}

%% file: biographies/hamila.tex
\begin{IEEEbiography}[{\includegraphics[width=1in,height=1.25in,clip,keepaspectratio]{hamila.jpg}}]{Prof. Ridha Hamila} received the MSc, LicTech with distinction, and PhD degrees from Tampere University of Technology (TUT), Tampere, Finland, in 1996, 1999, and 2002, respectively. Dr. Hamila currently a Full Professor at the Department of Electrical Engineering, Qatar University, Qatar. From 1994 to 2002 he held various research and teaching positions at TUT within the Department of Information Technology, Finland. From 2002 to 2003 he was a System Specialist at Nokia research Center and Nokia Networks, Helsinki. From 2004 to 2009 he was with Emirates Telecommunications Corporation, UAE. Also, from 2004 to 2013 he was adjunct Professor at the Department of Communications Engineering, TUT. His current research interests include mobile and broadband wireless communication systems, Mobile Edge Computing, Internet of Everything, and Machine Learning. In these areas, he has published over 200 journal and conference papers most of them in the peered reviewed IEEE publications, filed seven US patents, and wrote numerous confidential industrial research reports.  Dr. Hamila has been involved in several past and current industrial projects, Ooreedo, Qatar National Research Fund, Finnish Academy projects, EU research and education programs. He supervised a large number of under/graduate students and postdoctoral fellows. He organized many international workshops and conferences. He is a Senior Member of IEEE.
\end{IEEEbiography}

%% file: biographies/adel.tex
\begin{IEEEbiography}[{\includegraphics[width=1in,height=1.25in,clip,keepaspectratio]{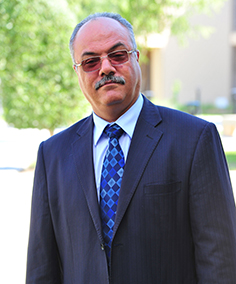}}]{Adel Gastli} (S'89, M'93, SM'00) received the B.Sc. degree in electrical engineering from the National Engineering School of Tunis, Tunisia, in 1985. He received the M.Sc. and the Ph.D. degrees from Nagoya Institute of Technology, Japan, in 1990 and 1993, respectively.,His professional career started with the National Institute for Standards and Intellectual Property, Tunis, Tunisia, where he worked for two years on the standardization and certification of electric products in Tunisia. Afterward, he joined the R\&D Department at Inazawa Works of Mitsubishi Electric Corporation in Japan and worked there until 1995. In August 1995, he joined the Electrical and Computer Engineering Department at Sultan Qaboos University, Muscat, Oman, where he served as the Head of the Department from September 2001 to August 2003 and from September 2007 to August 2009. He established the Renewable and Sustainable Energy Research Group (RASERG) at Sultan Qaboos University, in 2003, and was its Chair till January 2013. He also established the University Quality Assurance Office in February 2010 and served as its Director from February 2010 to January 2013. In February 2013, he joined the Electrical Engineering Department at Qatar University, Doha, Qatar, as a Professor and Kahramaa-Siemens Chair in energy efficiency. In August 2013, he was appointed the College of Engineering Associate Dean for Academic Affairs. In April 2014, he established the Qatar University Clean Energy \& Energy Efficiency Research Group (CE3RG) that he is currently chairing. He has authored/coauthored more than 150 papers in reputable journals and conferences. His research interests include energy efficiency, renewable energy, and smart grid.
\end{IEEEbiography}

%% file: biographies/serkan.tex
\begin{IEEEbiography}[{\includegraphics[width=1in,height=1.25in,clip,keepaspectratio]{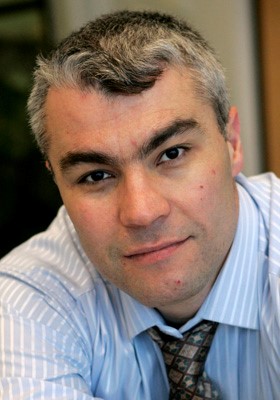}}]{Serkan Kiranyaz} was born in Turkey, 1972. He received his BS and MS degrees in Electrical and Electronics Department at Bilkent University, Ankara, Turkey, in 1994 and 1996, respectively. During 1996 – 2000 He worked as a Field Engineer in Schlumberger W\&T and Senior Researcher in Nokia Research Center, Tampere, Finland. He received his PhD degree in 2005 and his Docency at 2007 from Tampere University of Technology, Institute of Signal Processing respectively. He was working as a Professor in Signal Processing Department in the same university during 2009 to 2015 and he held the Research Director position for the department and also for the Center for Visual Decision Informatics (CVDI) in Finland. He currently works as a Professor in Qatar University, Doha, Qatar.
For more detailed information please refer to: \url{http://qufaculty.qu.edu.qa/mkiranyaz/}  and \url{https://www.researchgate.net/profile/Serkan_Kiranyaz} 
\end{IEEEbiography}

%% file: biographies/alemadi.tex
\begin{IEEEbiography}[{\includegraphics[width=1in,height=1.25in,clip,keepaspectratio]{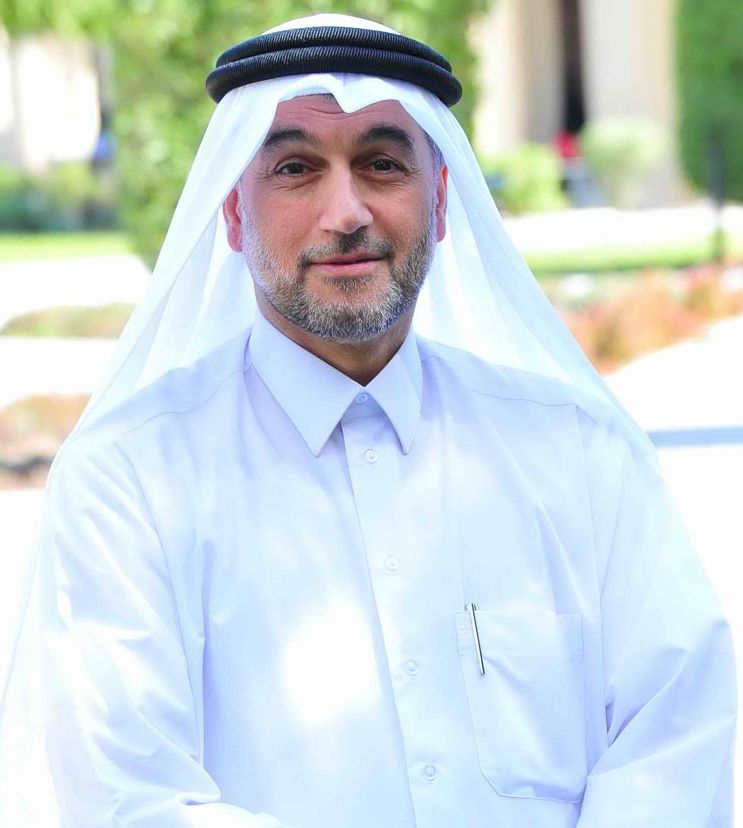}}] {Nasser Ahmed Al-Emadi}
received the B.S. and M.S. degrees, both in electrical engineering, from Western Michigan University, Kalamazoo, in 1989 and 1994, respectively, and the Ph.D. degree from Michigan State University, East Lansing, in 1999. He is currently a Full Professor and Chairperson of the Department of Electrical Engineering, Qatar University, Qatar. His research interests include operation, planning, and control of power systems, artificial neural networks (ANN) and machine learning.
\end{IEEEbiography}